\documentclass[twocolumn]{article}
\usepackage{graphicx}
\usepackage{amsmath,amssymb}
\usepackage{authblk}
\usepackage[margin=20truemm]{geometry}
\usepackage{float}
\usepackage{abstract}
\usepackage{cite}

\date{}
\title{\bf{Ferromagnetic diagonal stripe states in the two-dimensional Hubbard model with $U\lesssim\infty$}}  
\author[1]{\normalsize{Mitake Miyazaki}}
\author[2]{Takashi Yanagisawa}
\affil[1]{{\it Hakodate National College of Technology, 14-1, Tokura, Hakodate, 042-8501, Hokkaido, Japan}}
\affil[2]{\it{Electronics and Photonics Research Institute, National Institute of Advanced Industrial Science and Technology (AIST), Tsukuba Central 2, 1-1-1 Umezono, Tsukuba, 305-8568, Ibaraki, Japan}}

\begin{document}
\twocolumn[
\begin{@twocolumnfalse}
\maketitle
\vspace{-15mm}
\begin{abstract}
We have performed a variational Monte Carlo simulation to study the ground state of a two-dimensional Hubbard model on a square lattice in the strong coupling region. The energy gain of possible inhomogeneous electron states are computed as a function of $U$ when the hole density $\epsilon=1/8$ and next nearest-neighbor hopping $t'/t=-0.30$. The bond-centered ferromagnetic diagonal stripe state is stabilized in the strong coupling region ($U/t\geq$16), which is due to the gain of both kinetic energy and on-site Coulomb interaction energy due to the holon moving over the ferromagnetic domain and the gain of kinetic-exchange-interaction energy at the antiferromagnetic domain wall.\\ \ \\
\end{abstract}
\end{@twocolumnfalse}
]
\section{Introduction}
The ground state of the simple two-dimensional square lattice Hubbard model at half-filling is an antiferromagnetic (AF) Mott insulator for any finite positive value of $U/t$ (for $t'=0$). At $U=\infty$, doping only one hole drastically changes the AF insulator state to a ferromagnetic (FM) metallic state. This state is known as Nagaoka ferromagnetism~\cite{Nagaoka} which is driven by the motion of a hole. It is very interesting to investigate the electronic states that appear when the number of holes is increased or $U$ is decreased in this system. In the $U$ and finite-doping region, many numerical and theoretical studies have been conducted to determine whether fully polarized FM state persists~\cite{Becca}\cite{Carleo}, or whether partially polarized states (e.g., magnetic polarons~\cite{White1}) due to competing Nagaoka ferromagnetism and AF exchange interactions are stable, or whether other states such as spin-spiral states ~\cite{Duo,Kusakabe,Igoshev} and phase-separation replace Nagaoka ferromagnetism as ground states.

On the other hand, we focus our attention around the 1/8-doping in the middle $U$ case. The Hartree-Fock calculations~\cite{Poilblanc,Zaanen,Machida,Schulz} and numerical simulations such as the density matrix renormalization group~\cite{White2}, the constrained-path auxiliary-field quantum Monte Carlo~\cite{Chang}, the variational Monte Carlo (VMC)~\cite{Giamarch,Ido,Tocchio1}, etc.~\cite{Zheng,Qin} indicated the existence of stripe order with spatially non-uniform spin and charge density. This is a state in which holes doped in the AF Mott insulator do not disrupt the AF order and are arranged one-dimensionally on the domain wall between the two AF domains, forming incommensurate electronic states. In fact, the existence of the stripe-ordered state has been confirmed by neutron scattering experiments in copper oxide superconductors~\cite{Tranquada,Enoki}, for which the two-dimensional Hubbard model is considered to be a valid model. The formation mechanism of the stripe state is thought to be due to the kinetic energy gain of the hole on the domain wall and the antiferromagnetic exchange interaction gain on the AF domain.  Similarly, it has been discussed that superconductivity in strongly correlated cuprate superconductors may be driven by a gain in hole kinetic energy, unlike the usual BCS-type superconductivity~\cite{Feng,Wrobel,Maier,Ogata,Guo,Gull,Tocchio2,Yanagisawa1}. The kinetic energy gain is thought to play an important role in strongly correlated systems.

Now, it is a very interesting question to investigate what kind of electronic state can be changed from a fully polarized FM state with $U=\infty$ with a hole in the AF Mott insulator to the stripe states with $U\sim W$ ($W$ : band width) and 1/8 doping. Therefore, we use the VMC method that allows us to calculate a wide range of ground states from weakly correlated to strongly correlated states in order to search for stable ground states. As a clue, we investigate the energetic stability of the stripe states in the strongly correlated region.

This paper is organized as follows. Section 2 presents the model Hamiltonian and the trial wave function in variational Monte Carlo calculations. In Section 3, we present the energy difference between the normal and stripe states and spin and charge distributions for the striped wave function. In Section 4, we summarize and discuss the results obtained for FM diagonal stripe states.

\section{Model and Method}
The present study was undertaken to determine possible non-uniform charge distribution states in the two-dimensional (2D) $t$-$t'$-$U$ Hubbard model, 
\begin{eqnarray}
\hat{H}&=&\hat{H}_{\rm k}+\hat{H}_U\nonumber\\&=&-\sum_{i,j,\sigma }t_{ij}(\hat{c}_{i\sigma}^{\dagger}\hat{c}_{j\sigma}+{\rm h.c.})
+U\sum_i\hat{n}_{i\uparrow}\hat{n}_{i\downarrow},\ \ \   \label{hamil}
\end{eqnarray}
where the transfer energy $t_{ij}=t$, $t'$, and $0$, if sites $i$ and $j$ are nearest-, next-nearest neighbor and otherwise, respectively. In the following, we consider $t$ as the unit of energy. $\hat{c}_{i\sigma}^{\dagger}$ ($\hat{c}_{i\sigma}$) is the creation (annihilation) operator of the electron with  spin $\sigma$ ($\uparrow$ or $\downarrow$) at site $i$ ($i=1\sim N_{\rm site}$) and $\hat{n}_{i\sigma}=\hat{c}^{\dagger}_{i\sigma}\hat{c}_{i\sigma}$. $U$ is the on-site Coulomb interaction energy. 

In the VMC calculation, the variational energy is written as, $E_{\rm var}={\langle \Psi |\hat{H}|\Psi\rangle}/{\langle\Psi |\Psi\rangle}$. The trial wave function $|\Psi\rangle$ is defined by
$|\Psi\rangle=\hat{P}_{N_e}\hat{P}_{\rm G}\hat{P}_{\rm J}\hat{P}_{\rm DH}|\phi_{\rm MF}\rangle$, where $\hat{P}_{N_{\rm e}}$ is a projection operator that extracts only the components with a fixed total number of electrons $N_{\rm e}$. $\hat{P}_{\rm G}$ is the Gutzwiller projection operator given by $\hat{P}_{\rm G}=\prod_i(1-(1-g)\hat{n}_{i\uparrow}\hat{n}_{i\downarrow})$, where $g$ is a variational parameter in the range from $0$ to unity, which controls the on-site electron correlation. $\hat{P}_{\rm J}$ is the Jastrow-type projection operator $\hat{P}_{\rm J}=\prod_{\langle ij\rangle}h^{\hat{n}_{i}\hat{n}_{j}}$, which allows the occupancy of the nearest-neighbor sites to be modified by adjusting $h$ in the neighborhood of 1. $\hat{P}_{\rm DH}=\prod_i(1-(1-\eta)\prod_{\tau} \hat{d}_i (1-\hat{e}_{i+\tau}))$ is a correlation factor~\cite{Yokoyama1} between the double occupied site (doublon) and empty site (holon) on nearest-neighbors, where $0\leq\eta\leq 1$, $\hat{d}_i=\hat{n}_{i\uparrow}\hat{n}_{i\downarrow}$, $\hat{e}_i=(1-\hat{n}_{i\uparrow})(1-\hat{n}_{i\downarrow})$, and $\tau$ runs over all nearest-neighbor sites. $|\phi_{\rm MF}\rangle$ is a mean field (MF) wavefunction for non-uniform electron states, which is obtained from the following MF Hamiltonian,
\begin{eqnarray}
 \hat{H}_{\rm MF}=-\sum_{ij\sigma}t_{ij}\hat{c}_{i\sigma}^{\dagger}\hat{c}_{j\sigma}+\frac{U}{2}\sum_{i\sigma}\left( \rho_i-{\rm sgn}(\sigma)m_i\right) \hat{n}_{i\sigma}. 
\end{eqnarray}
where $t_{ij}$ is defined as that in eq. (1). The charge modulation $\rho_i$ and the spin modulation $m_i$ are described as
\begin{eqnarray}
 \rho_i &=& \rho\cos(\bf{q}\cdot (\bf{r}_i-{\bf r_0})), \\
 m_i &=& m\sin(\bf{Q}\cdot (\bf{r}_i-{\bf r_0})),
\end{eqnarray}
where $\rho$ and $m$ are variational parameters. The charge and spin configuration are characterized by incommensurate wave vectors ${\bf q}$ and ${\bf Q}$, respectively.  ${\bf Q}_{\rm VS}^{\rm AF}=(\pi\pm2\pi\delta, \pi)$ produces the AF vertical stripe state in which two adjacent AF magnetic domains are separated by a one-dimensional domain wall along the $y$-direction, resulting in a $\pi$-phase shift between the domain walls. $\delta$ is the incommensurability defined as the inverse of the period of the spin-stripe in the $x$-direction. The charge modulation period is fixed to half of the spin modulation period; ${\bf q}=2{\bf Q}$, where we deal only with stripes with even wavelength. ${\bf r}_0$ denotes the position of the domain boundary; ${\bf r}_0^{\rm SC}=(0, 0)$ corresponds to the site-centered (SC) type and  ${\bf r}_0^{\rm VBC}=(1/2,0)$ corresponds to the vertical bond-centered (BC) type. 

The AF diagonal stripe state, where the domain wall appears diagonally on the lattice, is represented by ${\bf Q}_{\rm DS}^{\rm AF}$ ($=(\pi\pm 2\pi\delta,\pi\pm 2\pi\delta)$). When the domain boundary is located at the center of the diagonal bond, ${\bf r}_0^{\rm DBC}=(1/2,1/2)$, the phase shifts on both sides, resulting in FM spin configurations on both sides of the domain boundary (Fig. 1(a)). Therefore, the AF diagonal stripe state is expected to gain the kinetic energy from the free movement of holes at the FM domain walls. As shown in Fig. 1(b), we assume a BC-FM diagonal stripe state with a diagonal wave vector ${\bf Q}_{\rm DS}^{\rm FM}$ ($=(\pm 2\pi\delta,\pm 2\pi\delta)$) and ${\bf r}_0^{\rm DBC}$, which consist of the diagonal FM magnetic domain and the diagonal AF domain boundary. (Similarly, a FM vertical stripe state with ${\bf Q}_{\rm VS}^{\rm FM}$ ($=(\pm 2\pi\delta,0)$)  can also be considered.) This configuration is advantageous for free moving holes to gain kinetic energy compared to Fig. 1(a). Furthermore, the energy gain due to the kinetic energy exchange interaction also occurs at the AF domain walls. 

\begin{figure}[H]
	\centering
		\includegraphics[scale=.33]{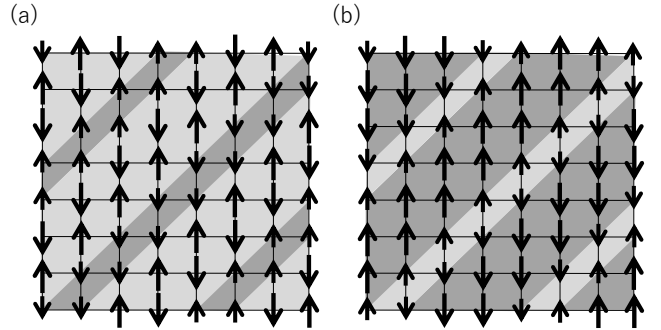}
	  \caption{Typical diagonal striped non-uniform electron configurations studied in this paper. The length of arrows is proportional to the spin-density.  These patterns have period-8 on both axes with regard to the spin. (a) Antiferromagnetic (AF) diagonal stripe state; the dark gray areas indicate ferromagnetic (FM) antiphase domain walls between AF diagonal striped domains. (b) FM diagonal stripe state; the light gray areas denote AF antiphase domain walls between FM diagonal striped domains.}\label{fig1}
\end{figure}

The energy expectation by the VMC method is optimized for a total number of Monte Carlo steps greater than $3\times 10^7$. To calculate the variational energy of the period-$1/\delta$ diagonal stripe states as shown in Fig. 1, the commensurability with $\delta$ is needed to guarantee the spin-periodicity along both $x$- and $y$-direction; the calculations are performed on a square lattice $L\times L$ ($L=8, 16$ and $24$). The periodic boundary conditions are applied in both directions. The hole density $\epsilon$ is fixed at 1/8-doping, namely $\epsilon=1-N_{\rm e}/N_{\rm s}=1/8$, where $N_{\rm s}$ is total number of sites. The stripe state satisfying the relations $\delta= \epsilon$ and $\delta= \epsilon/2$,  (i.e. period-8 and period-16 spin-stripe state, respectively) is assumed to examine the effective domain region when $U$ is large. We assume $t'=-0.30$ being suitable for cuprates~\cite{Pavarini} such as Bi-2201 in which the stripe state has been observed. It is thought that the next-nearest hopping $t'$ induces disorder in the strongly correlated system and produces inhomogeneous electronic states. Numerical works show that the next-nearest neighbor hopping term makes the stripe state more stable~\cite{Miyazaki,Ponsioen} and no phase separation appears in the 1/8-doped region~\cite{Yanagisawa2}. 

\section{Results}
\subsection{Energy difference between Normal and Stripe states}\label{4}
In Fig. 2, the energy difference per site between the normal and stripe states, $\Delta E=(E_{\rm normal}-E_{\rm stripe})/N_{\rm s}$, of various stripe states at 1/8-doping is shown as a function of $U$. The calculations were performed on the $16\times 16$ sites. The results show that FM stripe states (open symbols) are more energetically stable than AF-stripe states (filled symbols) at $U\geq16$. The stripe energy gain of the site-centered AF-vertical stripe state increases from weak-correlated region toward $U\sim10$ and decreases with increasing $U$ as the AF correlation ($\sim O(t^2/U$)) becomes weaker. On the other hand, the stripe energy gain of the period-8 bond-centered FM diagonal stripe state increases with increasing $U$, and the period-16 bond-centered FM diagonal stripe state is the most stable at $U\sim\infty$. Although the bond-centered AF-diagonal stripe state with period-8 behaves similarly with FM domain wall, the stripe energy gain is smaller than that for the FM diagonal stripe state. In addition, the bond-centered FM vertical stripe state (not shown in the Fig. 1) also has finite stripe energy gain in the strong correlation region. However, the FM diagonal stripe state has a larger the energy gain than the FM vertical stripe state, because the kinetic energy gain at the domain boundary is greater for the diagonal configuration than that for the vertical one. 

The stripe energy gain of the FM stripe state is not significantly affected by the setting of boundary conditions. It also decreases with decreasing $t'$, but finite values are obtained even at $t'=0$. The system size dependence of the energy gain per site of bond-centered FM diagonal stripe state with period-8 at $U=100$ is shown in inset of Fig. 1, where the energy gain does not decrease with increasing system size within statistic error. There is likely to be no size dependence for period-16 FM diagonal stripe as well as that for period-8 one. It should be noted that the fully polarized FM state is not stabilized in our system with $U\sim\infty$ unlike previous variational QMC results [2]. We expect this state to be sensitive to the lattice structure. 

\begin{figure}[H]
	\centering
		\includegraphics[scale=.39]{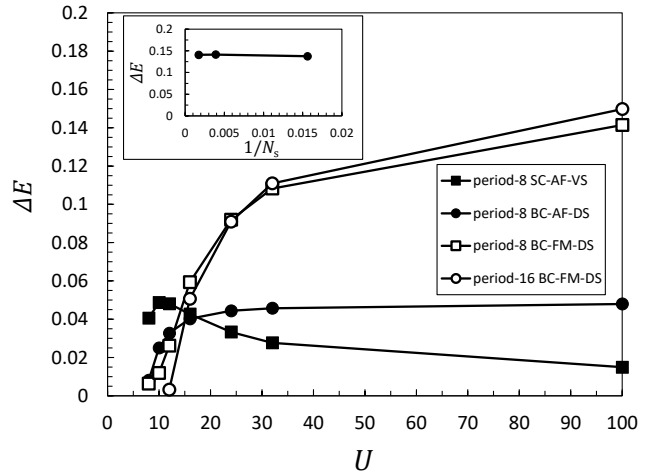}
	  \caption{Optimized energy difference per site between the normal and stripe states $\Delta E=(E_{\rm normal}-E_{\rm stripe})/N_{\rm s}$ as a function of $U$. The data is from $16\times 16$ lattices where electron number $N_e=224$ (i.e., 1/8-doping) and $t'=-0.30$. Filled squares indicate the period-8 site-centered AF vertical stripe state (8-period SC-AF-VS). Filled circles denote the period-8 bond-centered AF diagonal stripe state (8-period BC-AF-DS). Open squares (circles) denote the bond-centered FM diagonal stripe state (BC-FM-DS) with period-8 (period-16). Inset figure shows the system-size dependence of the optimized energy difference per site for the period-8 BC-FM-DS state with $U=100$ at 1/8-doping. The error bars are smaller than the size of symbols.}\label{fig2}
\end{figure}

\subsection{Three components of Energy difference}
In the following, to discuss the relative stability of the three states (the site-centered AF vertical stripe state, the bond-centered AF diagonal stripe state and the bond-centered FM diagonal stripe state) with period-8, we examine the stripe energy gain separately with contributions from the Coulomb interaction energy $E_U=\langle\hat{H}_U\rangle$ and the kinetic energy $E_{\rm k}=\langle\hat{H}_{\rm k}\rangle$. In addition, as illustrated in Fig. 3(a), $E_{\rm k}$ is divided into contributions from processes in which the total number of double occupied sites changes  (left panel) and does not change (right panel) when electrons hop from $i$ site to $j$ site.~\cite{Tocchio3} The former represents the intermediate process ($\sim O(t/U)$) of kinetic exchange interaction , and the latter expresses the motion of holes or doublons. In other words, the former provides more kinetic energy in the AF spin background and the latter in the FM one; $E_{\rm k}=E_{\rm k}^{\rm AF}+E_{\rm k}^{\rm FM}$. 

In Fig. 3(b), (c) and (d), we show the three components, $\Delta E_{\rm k}^{\rm AF}/N_{\rm s}$ (filled circles), $\Delta E_{U}/N_{\rm s}$ (open circles) and $\Delta E_{\rm k}^{\rm FM}/N_{\rm s}$ (filled squares) of the energy difference per site between the normal and stripe states as a function of $U$. Figure 3(b) for the period-8 site-centered AF vertical stripe state indicates that $\Delta E_{\rm k}^{\rm AF}/N_{\rm s}$ increases with increasing $U$ up to $U=12$ and then decreases (with respect to $\sim O(t/U)$). $\Delta E_{U}/N_{\rm s}$ changes from positive to negative at $U\sim 11$ and then approaches zero. Conversely, $\Delta E_{\rm k}^{\rm FM}/N_{\rm s}$ goes from negative to positive and then has a constant value. Thus, $U\sim 11$ is considered to be the turning point from a weakly correlated system to a strongly correlated system as $U$ increases. The stripe states are stabilized by interaction energy in the weakly correlated region and by kinetic energy gain in the strongly correlated region. These behaviors are similar to the results shown by Yokoyama et al.~\cite{Yokoyama2} in the superconducting state and uniform AF states, except that $\Delta E_{\rm k}^{\rm FM}$ changes to a positive value (This difference is due to the stripe state earning more the kinetic energy on the domain wall).

\begin{figure}[H]
	\centering
		\includegraphics[scale=.35]{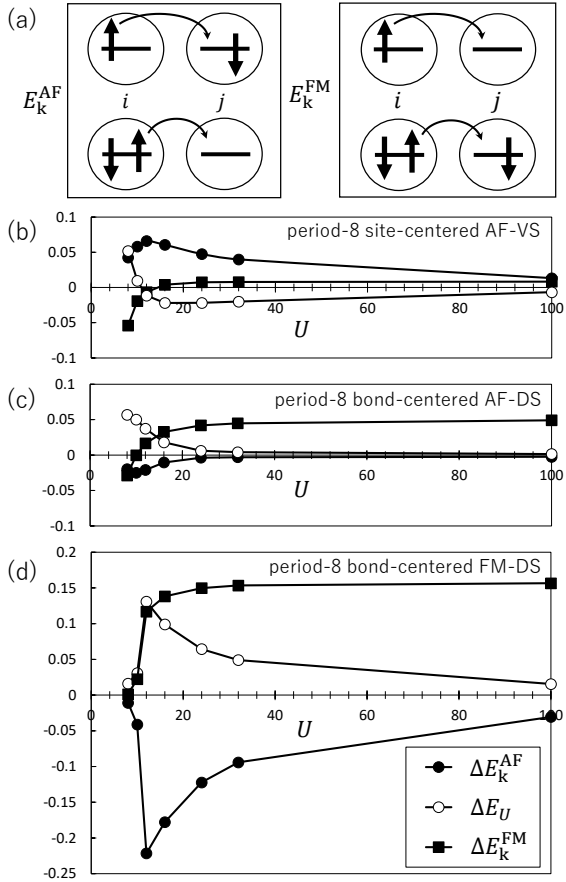}
	  \caption{(a) Illustration of an electron hopping from site $i$ to site $j$. The left (right) panel: The hopping with (without) changing the number of double occupied sites contributing to the kinetic energy part, $E^{\rm AF}_{\rm k}$ ($E^{\rm FM}_{\rm k}$). (b) The 3-components of energy difference between the normal state and the period-8 site-centered AF-VS, $\Delta E_{\rm k}^{\rm FM}$ (filled squares), $\Delta E_{\rm k}^{\rm AF}$ (filled circles) and $\Delta E_U$ (open circles),  as a function of $U$. (c) and (d) are same as (b) but for the period-8 bond-centered AFDS and for the period-8 bond-centered FMDS, respectively. }\label{fig3}
\end{figure}

For the bond-centered AF diagonal stripe state (Fig. 3(c)), $\Delta E_{U}/N_{\rm s}$ is positive; due to FM arrangement on the domain wall, the double occupancy probability is smaller than in the normal state, energetically more favorable than in the normal state. As $U$ increases, the total number of doublon number of double occupied sites decreases and $\Delta E_{U}/N_{\rm s}$ approaches zero. On the other hand, $\Delta E_{\rm k}^{\rm AF}/N_{\rm s}$ is almost zero because the gain on the AF domain cancels the loss on the FM domain wall. $\Delta E_{\rm k}^{\rm FM}/N_{\rm s}$ has a $U$-independent kinetic energy contribution on the FM domain. These features are more pronounced in the bond-centered FM diagonal stripe state as shown in Fig. 3(d). Although the loss of kinetic exchange interaction energy is large, but decreases with increasing $U$. $\Delta E_{\rm k}^{\rm FM}/N_{\rm s}$ and $\Delta E_{U}/N_{\rm s}$ are largest for the FM diagonal stripe state with FM domains. Therefore, in the strong coupling region, the bond-centered FM stripe state with extended FM domain are further stabilized, as shown in Fig. 2.

\subsection{Staggered magnetization, Charge density and Doublon density}

Here, we compare the optimized spatial electron distribution for the period-8 site-centered AF vertical stripe state at $U=16$, the period-8 bond-centered FM diagonal stripe state at $U=16$ and the period-16 bond-centered FM diagonal stripe state at $U=100$.  In Fig.4, the expectation values of the staggered magnetization, $(-1)^{x_i+y_i}\langle \hat{n}_{i\uparrow}-\hat{n}_{i\downarrow}\rangle$ for AF stripe state ($\langle \hat{n}_{i\uparrow}-\hat{n}_{i\downarrow}\rangle$ for FM stripe state), charge density, $\langle \hat{n}_{i\uparrow}+\hat{n}_{i\downarrow}\rangle$, and doublon density multiplied by $U$, $U\langle \hat{d}_{i}\rangle$, are denoted by filled squares, open circles,  and filled circles, respectively. In the period-8 site-centered AF vertical stripe state (Fig. 4(a)), holes are concentrated at the domain boundary because $U\langle \hat{d}_{i}\rangle$ is highest at the domain boundary and acts as more repulsive than the normal state. Aa a result, the kinetic energy contribution along the domain wall is the largest. There is also a finite doublon density inside the AF domain, which contributes to the energy gain of the kinetic exchange interaction.

On the contrary, as shown in Fig. 4(b), in the bond-centered FM diagonal stripe state, holes are distributed more on the FM domain and contribute to the kinetic energy gain. On the FM domain, the $U\langle \hat{d}_{i}\rangle$ is almost zero and little kinetic exchange interaction energy is gained, but it is highest at the domain boundary, where the kinetic exchange interaction energy gain can be obtained. At $U=100$ (Fig. 4(c)), the period-16 bond-centered FM diagonal stripe state has the same properties as the period-8 one. However, the doublon density decreases significantly in the strong correlation limit. A slight periodic modulation of the hole distribution is also observed inside the FM domain. The staggered magnetization of the FM stripe states has large values even near the domain boundary as shown in Figs. 3(b) and (c), and can be regarded as a twisted state of fully polarized FM state than the incommensurate spin-density-wave.

\begin{figure}[H]
	\centering
		\includegraphics[scale=.40]{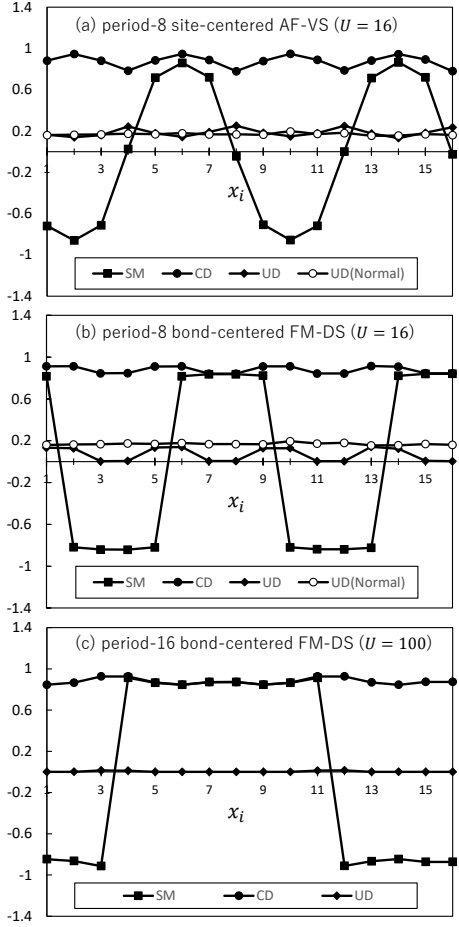}
	  \caption{Profiles of the expectation values of the staggered magnetization, (SM), the charge density, (CD), and the doublon density multiplied by $U$, (UD), along the $x_i$ direction at 1/8-doping are plotted by filled squares, filled circles and filled diamonds, respectively. (a) the period-8 bond-centered AF diagonal stripe state with $U=16$, (b) the period-8 bond-centered FM diagonal stripe state with $U=16$ and (c) with $U=100$. (b) and (c) are the most energetically stable cases at each $U$, as shown in Fig. 2. As a reference, UD for the normal state are plotted by open circles in (a) and (b).}\label{fig4}
\end{figure}

\section{Summary and Discussion}

The stripe energy gain of inhomogeneous electron distribution states in the two-dimensional square lattice Hubbard model with large $U$ are investigated using the variational Monte Carlo method. The system parameters were chosen to be $t'=-0.30$ and $1/8$-doping. We find that the gain in the kinetic exchange interaction energy decreases with increasing $U$, while in the large $U$ region, the gain in the kinetic energy due to the transfer of holes on the ferromagnetic domain region and the gain in the on-site Coulomb interaction energy exceed the loss in the kinetic exchange interaction energy, forming an incommensurate electron state with the ferromagnetic domain. 

One possible explanation as to why large kinetic energies of holes can be obtained in incommensurate striped ordered states is discussed below. The wave function in equation (2) can be regarded as the wave function of the twisted spin state with an extra factor $\sin(\bf{Q}^{\rm FM}\cdot(\bf{r}_i-\bf{r}_0))$ added to the modulated fully polarized FM state rather than that of the incommensurate density wave. Without this phase factor, the FM state with finite $S_z$ is not stabilized. As discussed by Dou\c{c}ot and Wen [5], we think that the increase in internal frustration due to finite hole doping into the strongly correlated systems is eliminated by twisting the spin state. They considered a spin-spiral state with a wavelength of system size $L$, whereas in the present work we assume the incommensurate spin density wave with a wavelength $1/\delta$. As shown in Figure 2, the period of twists at $U\sim\infty$ is $\delta=\epsilon/2$ and seems to correlate with the hole concentration causing frustration. As $U$ decreases, the bond centered FM diagonal stripe state with the relation of $\delta=\epsilon$ becomes more stable than that with the relationship $\delta=\epsilon/2$. This appears to be due to the enhancement of AF correlations with decreasing $U$ and the kinetic exchange energy is gained by shortening the twisting period. 

There may be some relationship between the formation of stripes in the strongly coupled region and that in the moderate coupling region, since the frustration due to fermionic statistics of doped holes is a common underlying problem. However, it is an open question as to whether the twist of the spin wavefunction eliminates the frustration in the stripe state in the moderate coupling region.

\section*{Acknowledgements}
This work was partly achieved through the use of Hokkaido University High-Performance Intercloud at the information initiative center, Hokkaido University, Sapporo, Japan.

\end{document}